# Comment: Citation Statistics

**Sune Lehmann, Benny E. Lautrup and Andrew D. Jackson**

*Abstract.* We discuss the paper "Citation Statistics" by the Joint Committee on Quantitative Assessment of Research. In particular, we focus on a necessary feature of "good" measures for ranking scientific authors: that good measures must able to accurately distinguish between authors.

## 1. INTRODUCTION

The Joint Committee on Quantitative Assessment of Research (the Committee) has written a highly readable and well argued report discussing common misuses of citation data. The Committee argues convincingly that even the meaning of the "atom" of citation analysis, the citation of a single paper, is nontrivial and not easily converted to a measure of research quality. The Committee also emphasizes that the assessment of research based on citation statistics always reduces to the creation of *ranked lists* of papers, people, journals, etc. In order to create such a ranking of scientific authors, it is necessary to describe each author's full publication and citation record to a single scalar measure, $\mathcal{M}$. It is obvious that any choice of $\mathcal{M}$ that is independent of the citation record (e.g., the number of papers published per year) is likely to be a poor measure of research quality. However, it is less clear what constitutes a "good" measure of an author's full citation record. We have previously discussed this question in some detail [1, 2], but in the light of the present report, the subject appears to merit further discussion. Below, we elaborate on the definition of the term "good" in the context of ranking scientific authors by describing how to assign objective (i.e., purely statistical) uncertainties to any given choice of $\mathcal{M}$.

## 2. IMPROBABLE AUTHORS

It is possible to divide the question of what constitutes a "good" scalar measure of author performance into two components. One aspect is wholly subjective and not amenable to quantitative investigation. We illustrate this with an example. Consider two authors, $A$ and $B$, who have written 10 papers each. Author $A$ has written 10 papers with 100 citations each and author $B$ has written one paper with 1000 citations and 9 papers with 0 citations. First, we consider an argument for concluding that author $A$ is the "better" of the two.

In spite of varying citation habits in different fields of science, the distribution of citations within each field is a highly skewed power-law type distribution (e.g. see [3, 4]). Because of the power-law structure of citation distributions, the citation record of author $A$ is more improbable than that of author $B$. It is illuminating to quantify the difference between the two authors using a real dataset. Here, we use data from the SPIRES database for high energy physics (see [3] for details regarding this dataset). The *citation summary* option on the SPIRES website returns the number of papers for a given author with citations in each of six intervals. These intervals and the probabilities that papers will fall in these bins are given in Table 1. The probability, $P(\{n_i\})$, that an author's actual citation record of $N$ papers was obtained from a random draw on the citation distribution $P(i)$ is readily calculated by multiplying the probabilities of drawing the author's number of publications in the different categories, $n_i$, and correcting for the number of permutations.

*Sune Lehmann is Postdoctoral Researcher, Center for Complex Network Research, Department of Physics, Northeastern University, Boston, Massachusetts, USA and Center for Cancer Systems Biology, Dana-Farber Cancer Institute, Harvard University, Boston, Massachusetts, USA e-mail: lehmann@neu.edu. Benny E. Lautrup is Professor of Theoretical Physics, The Niels Bohr Institute, University of Copenhagen, Copenhagen, Denmark. Andrew D. Jackson is Professor of Physics, Niels Bohr International Academy, The Niels Bohr Institute, Copenhagen, Denmark.*

TABLE 1
*The search option* citation summary *at the SPIRES website returns the number of papers for a given author with citations in the intervals shown. The probabilities of getting citations in these are intervals are listed in the third column*

| Paper category | Citations | Probability $P(i)$ |
|---|---|---|
| Unknown papers | 0 | 0.267 |
| Less known papers | 1–9 | 0.444 |
| Known papers | 10–49 | 0.224 |
| Well-known papers | 50–99 | 0.0380 |
| Famous papers | 100–499 | 0.0250 |
| Renowned papers | 500+ | 0.00184 |

$$P(\{n_i\}) = N! \prod_i \frac{P(i)^{n_i}}{n_i!}.$$

If a total of $N$ papers is drawn at random on the citation distribution, the most probable result, $P(\{n_i\}_{\max})$, corresponds to $n_i = NP(i)$ papers in each bin. The quantity

$$r = -\log_{10}\left[\frac{P(\{n_i\})}{P(\{n_i\}_{\max})}\right],$$

is a useful measure of this probability which is relatively independent of the number of bins chosen. In the case of author $A$ we find the value $r_A = 14.4$, and for author $B$ we find $r_B = 5.33$. In spite of the fact that $A$ and $B$ have the same average number of citations, the record of author $B$ is roughly $10^9$ times more probable that that of author $A$! We do not claim that the "unlikelihood" measure $r$ captures the richness of the full data set nor that it captures the complexity of individual citation records,[1] but we do claim that the extreme improbability of author $A$ might convince some to choose her over author $B$.

On the other hand, if one believes that the most highly cited papers have a special importance—that they contain scientific results that are particularly significant or noteworthy—one might reasonably prefer $B$ over $A$. A famous proponent of this view is the father of bibliometrics, E. Garfield, who dubbed such papers "citation classics" [5]. No amount of quantitative research will convince a supporter of citation classics that the improbability of a citation record is a better measure of the scientific significance of an author or vice versa; this judgment is strictly subjective.

## 3. DISCRIMINATORY ABILITY

As mentioned above, scalar measures of author quality also contain an element that can be assessed objectively. Whatever the intrinsic and value-based merits of the measure, $\mathcal{M}$, assigned to every author, it will be of no practical value unless the corresponding uncertainty, $\delta\mathcal{M}$ is sufficiently small. From this point of view, the "best" choice of measure will be that which provides maximal discrimination between scientists and hence the smallest value of $\delta\mathcal{M}$. If a measure cannot be assigned to a given author with suitable precision, the subjective issue of its relation to author quality is rendered moot. Below we outline how the question of deciding which of several proposed measures is most discriminating, and therefore "best," can be addressed quantitatively using standard statistical methods.

The model that authors $A$ and $B$ draw their citation records on the total citation distribution $P(i)$ is quite primitive. This is indicated by the fact that the numerical values of $r$ for both $A$ and $B$ are uncomfortably large. It is more reasonable to assume that each author's record was drawn on some sub-distribution of citations. By using various measures of author quality to construct such sub-distributions, we can gauge their discriminatory abilities. We formalize this idea below.

We start by binning all authors according to some tentative indicator, $\mathcal{M}$, obtained from their full citation record. The probability that an author will lie in bin $\alpha$ is denoted $p(\alpha)$. Similarly, we bin each paper according to the total number of its citations.[2] The full citation record for an author is simply the set $\{n_i\}$. For each author bin, $\alpha$, we then empirically construct the conditional probability distribution, $P(i|\alpha)$, that a single paper by an author in bin $\alpha$ will lie in citation bin $i$. These conditional probabilities are the central ingredient in our analysis. They can be used to calculate the probability, $P(\{n_i\}|\alpha)$, that any full citation record was actually drawn at random on the conditional distribution, $P(i|\alpha)$ appropriate for a fixed author bin, $\alpha$. Bayes' theorem allows us to invert this probability to yield

$$(1) \qquad P(\alpha|\{n_i\}) \sim P(\{n_i\}|\alpha)p(\alpha),$$

where $P(\alpha|\{n_i\})$ is the probability that the citation record $\{n_i\}$ was drawn at random from author bin $\alpha$. By considering the actual citation histories of authors in bin $\beta$, we can thus construct the probability $P(\alpha|\beta)$, that the citation record of an author initially assigned to bin $\beta$ was drawn on the distribution appropriate for bin $\alpha$. In other words, we can determine the probability that an author assigned to bin $\beta$ on the basis of the tentative indicator should actually be placed in bin $\alpha$. This allows us to determine both the accuracy of the initial author assignment and its uncertainty in a purely statistical fashion.

---

[1] For example, it is possible to be an improbably bad author.

[2] We use Greek letters when binning with respect to $\mathcal{M}$ and Roman for binning citations.

While a good choice of indicator will assign each author to the correct bin with high probability, this will not be the case for a poor measure. Consider extreme cases in which we elect to bin authors on the basis of indicators unrelated to scientific quality, e.g., by hair/eye color or alphabetically. For such indicators, $P(i|\alpha)$ and $P(\{n_i\}|\alpha)$ will be independent of $\alpha$, and $P(\alpha|\{n_i\})$ will be proportional to the prior distribution $p(\alpha)$. As a consequence, the proposed indicator will have no predictive power whatsoever. The utility of a given indicator (as indicated by the statistical accuracy with which a value can be assigned to any given author) will obviously be enhanced when the basic distributions $P(i|\alpha)$ depend strongly on $\alpha$. These differences can be formalized using the standard Kullback–Leibler divergence. The method outline above was applied to several measures of author performance in [1, 2]. Some familiar measures, including papers per year and the Hirsch index [6], do not reflect an author's full citation record and are little better than a random ranking of authors. The most accurate measures (e.g., mean or median citations per paper) are able to assign authors to the correct decile bin with 90% confidence on the basis of approximately 50 papers. Since the accuracy of assignment grows exponentially with the number of papers, the evaluation of authors with significantly fewer papers is not likely to be useful.

## 4. DATA HOMOGENEITY

The average number of citations for a scientific paper varies significantly from field to field. A study of the impact factors on *Web of Science* [7] show that an average paper in molecular biology and biochemistry receives approximately 6 times more citations than a paper in mathematics. Such distinction, which are unrelated to field size or publication frequency, are entirely due to differences in the accepted referencing practice which have emerged in separate scientific fields. It is obvious that a fair comparison of authors in different fields must recognize and correct for such cultural inhomogeneities in the data. This task is more difficult than might be expected since significant differences in referencing/citation practice can be found at a surprisingly microscopic level. Consider the following subfield hierarchy:

$$\begin{aligned} \text{physics} &\rightarrow \text{high energy physics} \\ &\rightarrow \text{high energy theory} \\ &\rightarrow \text{superstring theory.} \end{aligned}$$

Study of the SPIRES database reveals that the natural assumption of identical referencing/citation patterns for string and non-string theory papers is grossly incorrect. Since its emergence in the 1980s, string theory has evolved into a distinct and largely self-contained subfield with its own characteristic referencing practices. Specifically, our studies indicate that the average number of citations/references for string theory papers is now roughly twice that of non-string theory papers in theoretical high energy physics. Any attempt to compare string theorists with non-string theorists will be meaningless unless these non-homogeneities are recognized and taken into consideration. Unfortunately, such information is not usually supplied by or readily obtainable from commercial databases.

## 5. IN SUMMARY

The Committee's report provides a much needed criticism of common misuses of citation data. By attempting to separate issues that are amenable to statistical analysis from purely subjective issues, we hope to have shown that serious statistical analysis does have a place in a field that is currently dominated by ad hoc measures, rationalized by anecdotal examples and by comparisons with other ad hoc measures. The probabilistic methods outlined above permit meaningful comparison of scientists working distinct areas with minimal value judgments. It seems fair, for example, to declare equality between scientists in the same percentile of their peer groups. It is similarly possible to combine probabilities in order to assign a meaningful ranking to authors with publications in several disjoint areas. All that is required is knowledge of the conditional probabilities appropriate for each homogeneous subgroup.

We emphasize that meaningful statistical analysis requires the availability of data sets of *demonstrated* homogeneity. The common tacit assumption of homogeneity in the absence of evidence to the contrary is not tenable. Finally, we note that statistical analyses along the lines indicated here are capable of identifying groups of scientists with similar citation records in a manner which is both objective and of quantifiable accuracy. The interpretation of these citation records and their relationship to intrinsic scientific quality remains a subjective and value-based issue.

## REFERENCES

[1] LEHMANN, S., JACKSON, A. D. and LAUTRUP, B. E. (2006). Measures for measures. *Nature* **444** 1003.
[2] LEHMANN, S., JACKSON, A. D. and LAUTRUP, B. E. (2008). A quantitative analysis of indicators of scientific performance. *Scientometrics* **76** 369.
[3] LEHMANN, S., LAUTRUP, B. E. and JACKSON, A. D. (2003). Citation networks in high energy physics. *Phys. Rev. E* **68** 026113.
[4] REDNER, S. (1998). How popular is your paper? An empirical study of the citation distribution. *European Physics Journal B* **4** 131.
[5] GARFIELD, E. (1977). Introducing citation classics: The human side of scientific papers. *Current Contents* **1** 1.
[6] HIRSCH, J. E. (2005). An index to quantify an individual's scientific output. *Proc. Natl. Acad. Sci.* **102** 16569.
[7] THOMPSON SCIENTIFIC. Web of science, Address retrieved October 2008.
http://scientific.thomson.com/products/wos/.